\documentclass[twocolumn,fleqn,english,aps,prd,preprintnumbers,
showkeys,nofootinbib,superscriptaddress,floatfix,tightenlines]{revtex4-1} 
\usepackage[utf8]{inputenc}
\usepackage{amsmath}
\usepackage{amsthm}
\usepackage{amssymb}
\usepackage{graphicx}
\usepackage{amsfonts}
\usepackage{amscd}
\usepackage{amsxtra}
\usepackage{epstopdf}
\usepackage{color}
\usepackage{dcolumn}
\usepackage{bm}
\usepackage{multirow}

\makeatletter

\usepackage[normalem]{ulem}
\usepackage[dvipsnames]{xcolor}
\usepackage{array}
\usepackage{slashed}
\renewcommand{\sout}{\bgroup \color{red} \ULdepth=-.5ex \ULset}

\DeclareGraphicsExtensions{.jpg,.pdf,.png}
\newcommand{\Tr}{\mathrm{Tr}}

\begin{document}
\preprint{INHA-NTG-08/2021}
\title{Singly heavy baryons in nuclear matter from an SU(3) chiral
  soliton model}  
\author{Ho-Yeon Won}
\email{hoywon@inha.edu}
\affiliation{Department of Physics, Inha University, Incheon 22212,
   Korea} 

\author{Ulugbek Yakhshiev}
\email{yakhshiev@inha.ac.kr}
\affiliation{Department of Physics, Inha University, Incheon 22212,
  Korea}
\affiliation{Theoretical Physics Department, National University
of Uzbekistan, Tashkent 700174,
  Uzbekistan}

 \author{Hyun-Chul Kim}
\email{hchkim@inha.ac.kr}
\affiliation{Department of Physics, Inha University, Incheon 22212,
  Korea} 
\affiliation{School of Physics, Korea Institute for Advanced Study 
  (KIAS), Seoul 02455,  Korea}

\begin{abstract}
We investigate how the masses of the singly heavy baryons undergo
changes in nuclear matter, based on a medium-modified SU(3) chiral
soliton model. Having explained the bulk properties of nuclear matter,   
we discuss the masses of the singly heavy baryons
in nuclear matter. We generalize the vector-meson Lagrangian
including the heavy-meson soliton interaction.
The mass spectrum of the singly heavy baryon are obtained with the
effects of explicit SU(3) symmetry breaking considered as a
perturbation. The results show that the mass of the singly heavy
baryon mass in nuclear medium is rather sensitive to the medium
modifications of the heavy meson mass. 
\end{abstract}
\maketitle

\section{Introduction}
Understanding the medium modifications of light hadrons has been one of
the most important issues in hadronic and nuclear physics over
decades, since it sheds light on the fundamental features of quantum
chromodynamics such as the spontaneous breakdown of chiral symmetry 
and quark confinement (see following reviews~\cite{Drukarev:1991fs,
  Birse:1994cz, Brown:1995qt, Saito:2005rv, Hayano:2008vn}). It also
provides an important clue about how neutron stars are
formed~\cite{Lattimer:2006xb}, in particular, how the two solar-mass 
problem of the neutron star can be explained. Thus, it is essential to 
answer the questions as to how properties of the baryons 
undergo changes in nuclear medium to describe the neutron star. 
While medium modifications of heavy hadrons in nuclear matter have
been much less studied than those of light ones, there have been
several works~\cite{Dover:1977jw, Gatto:1978ka, Bando:1981ti,
  Gibson:1983zw} on charmed nuclei soon after the charmonium  
$J/\psi$ and $\Sigma_c$ were 
found~\cite{E598:1974sol,SLAC-SP-017:1974ind, 
  Cazzoli:1975et}. Then the heavy baryons in nuclear matter have been
sparsely studied~\cite{Tsushima:2002cc, Tsushima:2002sm, Wang:2011hta,  
  Wang:2011yj}. Recently, interest in heavy hadrons was renewed as the
experiments on them including exotic heavy hadrons have yielded
unprecedented findings~\cite{Choi:2003ue, LHCb:2015yax,LHCb:2020jpq,
  LHCb:2021ptx, LHCb:2021chn} (see a recent review for the status of
experiments on nonstandard heavy hadrons~\cite{Olsen:2017bmm}). This
has also triggered the investigation on heavy baryons in nuclear
matter~\cite{Ohtani:2017wdc, Azizi:2018dtb, Carames:2018xek,
  Yasui:2018sxz, Yasui:2019ogk, Haidenbauer:2020uci} (see also a
review~\cite{Hosaka:2016ypm}). Since there are no experimental data on
how the heavy baryons undergo changes in nuclear medium and in nuclei, 
it is of great importance to study the medium modification of them
theoretically to guide future experiments. 

In the present work, we investigate the in-medium modification of the 
singly heavy baryons based on an SU(3) chiral soliton model with a
bound-state approach~\cite{Momen:1993ax}. The model is distinguished
from a usual bound-state approach in which a heavy baryon emerges as a
bound state of a heavy meson and a baryon as a chiral
soliton~\cite{Callan:1985hy, Blaizot:1988ge, Rho:1990uy, Rho:1992yy,
  Weigel:1993zd}, regarding both the hyperons and kaons as 
heavy particles. In Ref.~\cite{Momen:1993ax}, on the other hand,
the singly heavy baryons can be constructed as a bound state of a
heavy meson and an SU(3) baryon as a soliton. This means that the
Wess-Zumino-Witten (WZW) term~\cite{Witten:1983tw} exists in the model
and constrains the allowed representations for the collective
Hamiltonian together with the kinetic term of the heavy mesons. What
is interesting is that the collective Hamiltonian describes the
bosonic soliton as if it had come from the $N_c-1$ valence
quarks~\cite{Yang:2016qdz}~\footnote{When $N_c$ is taken to be three,
  it corresponds to the light diquark. The present model bears a
  certain similarity to the pion mean-field approach for the singly
  heavy baryons~\cite{Yang:2016qdz, Kim:2018cxv}}. Then the singly 
heavy baryon emerges as a coupled state with the single heavy quark.
In addition, the model contains the vector-meson degrees of freedom,
which improves the semiclassical binding energy.  
This SU(3) chiral soliton model with a bound-state approach is 
simpler and clearer than those mentioned previously, so that 
it can be easily modified in medium. Since we have already constructed
the in-medium modified soliton model with vector-meson degrees of
freedom~\cite{Jung:2012sy}, it is rather straightforward to modify the
effective Lagrangian for this model. The only new ingredient is the
heavy-meson degrees of freedom, which was naturally introduced in such
a way that it complies with heavy-quark spin symmetry. Thus, we will
scrutinize how the in-medium modification of the heavy meson
influences the mass spectrum of the singly heavy baryons in nuclear
matter. This is a very interesting and remarkable feature of the
present approach, since it relates directly the medium modification
of the heavy mesons to the singly heavy baryons in the same manner as
the change of the vector mesons affects the nucleon and
$\Delta$ isobar in nuclear matter within the Skyrme models. 

We sketch the present work as follows: In Section II, we introduce the
medium functionals to modify the effective chiral Lagrangian. Since
the medium modification of the pion and the vector mesons was done in
Ref.~\cite{Jung:2012sy}, we follow the same method. We introduce an
additional parameter to explain how the heavy degrees of freedom
undergo changes in nuclear matter. This parameter governs the
dependence of the mass splitting of singly heavy baryons on the nonzero
nuclear density. In Section III, we show how the in-medium parameters
can be fixed by reproducing the bulk properties of nuclear matter such
as the volume energy and the pressure as functions of the nuclear 
density. In Section IV, we discuss the numerical results for the mass
spectra of the baryon antitriplet and sextet. Section V is devoted to
summary and conclusions of the present work. 
 
\section{Formalism}
In the present section we first formulate the medium modifications of
the effective chiral Lagrangian and SU(3) light baryons. Then we
proceed to implement the changes of the heavy mesons in nuclear
matter. 
\subsection{Light mesons in nuclear matter}
A medium modified Lagrangian of the system consisting of the light
pseudoscalar and vector mesons in nuclear matter  
within the SU(2) framework is given in Ref.\,\cite{Jung:2012sy}.  We
generalize the idea proposed in Ref.~\cite{Jung:2012sy}, which provides
a practical method to describe how the nucleon is modified in nuclear
medium, so that we can study the changes of properties of heavy
baryons. Thus, we reformulate the medium-modified Lagrangian starting
from the free-space Lagrangian presented in
Refs.~\cite{Gupta:1993kd,Momen:1993ax}.  Note that
Ref.\,\cite{Jung:2012sy} without medium modifications is in line with  
Refs.\,\cite{Gupta:1993kd,Momen:1993ax} in the flavor SU(2) case.
This means that all the methods developed in Ref.~\cite{Jung:2012sy}
can be easily generalized to the flavor SU(3) case, so that the
effective Lagrangian can be written as\footnote{Asterisks in the
superscript mean the explicit medium modification. For more details
of the Lagrangian within the SU(2) framework see
Ref.\,\cite{Jung:2012sy}. }  
\begin{align}
\label{Llight}
\mathcal{L}_{\rm
  light}^{*}&=\mathcal{L}_{\pi}^{*}+\mathcal{L}_{V}^{*} +
              \mathcal{L}_{\pi  V}^{*}+  \mathcal{L}_{\pi\omega}^*,\\
\label{Lpi}
\mathcal{L}_{\pi}^{*}&=-\frac{F_{\pi}^{*2}}{8}\mathrm{Tr}[\partial_{\mu}U
                       \partial_{\mu}U^{\dagger}],\\ 
\label{LV}
\mathcal{L}_{V}^{*}&=-\frac{1}{4}\mathrm{Tr}[F_{\mu\nu}^{*}(\rho)
                     F^*_{\mu\nu}(\rho)],\\ 
\label{LpiV}
\mathcal{L}_{\pi V}^{*}&=-\frac{m_v^{*2}}{2g^{*2}}\mathrm{Tr}
                         [(g^{*}\rho_{\mu}^*-v_{\mu}^+)^{2}], 
\end{align}
where $F_\pi^*$ is the medium-modified pion decay constant with its
free space value is $F_\pi=132$\,MeV.  $m_v^*$ and $g^*$ stand for the
medium-modified mass of the vector mesons and vector coupling
constants, respectively. Here the first three terms in
Eq.\,(\ref{Llight}) respectively describe pseudoscalar, vector and
pseudoscalar-vector interaction terms in nuclear matter in Euclidean
space. The first term $\mathcal{L}_{\pi}^{*}$, Eq.\,(\ref{Lpi})
denotes the well-known Weinberg kinetic term for the pseudoscalar
meson fields in a unitary form 
\begin{align}
U(\bm{r}) = \exp\left[i \lambda^a \pi^a \right] =  \xi^2,
\end{align}
where $\pi^a$ designate the pseudoscalar meson fields with $a=1,\cdots,
N_f^2-1$.  In the flavor SU(2), the pion fields are coupled to the
spatial coordinates, which is known as the hedgehog Ansatz or more
generally the hedgehog symmetry. It provides a minimal generalization
of spherical symmetry~\cite{Pauli:1942kwa}. So, the pion fields can be
expressed as
\begin{align}
  \label{eq:1}
\pi^a (\bm{r}) = n^a F(r)
\end{align}
with $n^a=x^a/r$ and $r=| \bm{r}|$. $F(r)$ is called the profile
function of the chiral soliton, which will be found by solving the
classical equations of motion for $\pi^a$ obtained from the effective
Lagrangian. In the case of $N_f=3$, the simplest generalization can
be acquired by the trivial embedding~\cite{Witten:1983tw}
\begin{align}
&\xi=\left(\begin{array}{ccc}
\exp[\frac{i}{2}(\hat{\bm{n}}\cdot\bm{\tau}F(r))] & 0\\
0 & 1\\
\end{array}\right),
\label{xi}
\end{align}
which preserves the hedgehog symmetry. $\bm{\tau}$ is the usual Pauli
matrices. 

The $3\times3$ field strength tensor in Eq.~\eqref{LV} is
expressed as 
\begin{align}
F^*_{\mu\nu}=\partial_\mu\rho_\nu^*-\partial_\nu\rho_\mu^*-i
  g^*[\rho_\mu^*,\rho_\nu^*], 
\end{align}
where 
\begin{align}
&\rho_{\mu}^*=\left(\begin{array}{ccc}
\frac{1}{\sqrt{2}}(\omega_{\mu}+\tau^a\rho^{*a}_{\mu}) & 0\\
0 & 1\\
\end{array}\right),\cr
&\rho_{i}^{*a}=\frac{\epsilon_{ika}\hat{n}^k}{\sqrt{2}g^*}
                         \frac{G(r)}{r}, \qquad \rho_{4}^{*a}=0,\cr 
&\omega_i=0, \qquad \omega_4=i\omega(r).
\label{vecc}
\end{align}
Here $G(r)$ and $\omega(r)$ denote respectively the profile
functions for the $\rho$ and $\omega$ mesons. The medium-modified
vector coupling $g^*$ takes the value $g=3.93$ in free space. 
The third term $\mathcal{L}_{\pi V}^{*}$ Eq.\,(\ref{LpiV})
describes the pseudoscalar-vector interactions, where the currents 
$v_\mu^\pm$ are written as 
\begin{align}
v_\mu^\pm=\frac{i}{2}(\xi\partial_\mu\xi^\dagger\pm\xi^\dagger\partial_\mu\xi)
\end{align}
and $m_v^*$ is taken to be the $\rho$ meson mass, i.e. $m_v=770$\,MeV
in free space. Finally, the last term $\mathcal{L}_{\pi\omega}^*$
represents the WZW term
\begin{align} 
\mathcal{L}_{\pi \omega}^{*}&=\frac{3}{\sqrt{2}}g^*\omega_{\mu}B_{\mu},
\end{align}
which couples the $\omega_\mu$ meson to the baryon current 
\begin{equation}
B^\mu=\frac{\epsilon^{\mu\nu\alpha\beta}}{24\pi^2}\Tr
\{(U^\dagger \partial_\nu U)(U^\dagger \partial_\alpha U)(U^\dagger
\partial_\beta U)\}. 
\end{equation}
The profile functions satisfy the following boundary conditions 
\begin{align}
F(0)=\pi,\quad G(0) =2, \quad \omega'(0)=0,\cr
  F(\infty)=G(\infty)=\omega(\infty)=0,
\end{align}
which yield the finite energy solutions. The details of the numerical
minimization process can be found in Ref.\,\cite{Jung:2012sy}. 

\subsection{Heavy mesons in nuclear matter}
We take the heavy meson part of the Lagrangian from
Ref.\,\cite{Momen:1993ax}. Note that the Lagrangian was constructed in
a way that heavy-quark flavor-spin symmetry is satisfied. We
modify the heavy mesons in the same manner as the light degrees of
freedom. Thus, the effective Lagrangian for the heavy mesons are
expressed as 
\begin{align}
\frac{\mathcal{L}_{\rm{heavy}}^*}{M}&=iV_{\mu}\mathrm{Tr}
 [H\big(\partial_{\mu}-\mathnormal{i}\alpha
 g^{*}\rho_{\mu}^{*}-\mathnormal{i}
 (1-\alpha)\mathnormal{v}_{\mu}^+\big)\bar{H}]\cr  
&+\mathnormal{i}d\,\mathrm{Tr}[H\gamma_{\mu}
\gamma_{5}v_{\mu}^-\bar{H}]\cr 
&+\frac{\mathnormal{i}c}{\mathnormal{m_{v}^{*}}}
  \mathrm{Tr}[H\gamma_{\mu}
  \gamma_{\nu}F_{\mu\nu}^{*}(\rho)\bar{H}],
\label{Lheavy}
\end{align}
where $H$ denotes the heavy super-field expressed by 
\begin{align}
H=\frac{1-i\gamma_{\mu}V_{\mu}}{2}(i\gamma_{5}P +
  i\gamma_{\nu}Q_{\nu}),\quad\bar{H}\equiv\gamma_{4}
  H^{\dagger}\gamma_{4}.
\label{heavysuperfield}
\end{align}
$P$ and $Q_\nu$ designate respectively the pseudoscalar and vector
heavy mesons. $V_\mu$ stands for the four-vector velocity of the heavy
quark, which imposes the superselection rule~\cite{Isgur:1990yhj,
  Georgi:1990um} on the heavy field. The values of the remaining
constants are given by $\alpha\simeq -2.9$, $d=0.53$ and $c=1.6$,
which are determined by using the experimental
data~\cite{Gupta:1993kd,PDG}.  

In the rest frame of the heavy meson $(V_{i}=0)$, the $4\times4$ heavy
superfield $\bar{H}$ in Eq.(\ref{heavysuperfield}) has only the 
nonvanishing elements in the lower-left $2\times2$ sub-block: 
\begin{align}
\bar{H}=\left(\begin{array}{ccc}
0 & 0 \\
\bar{H}_{lh}^{b} & 0 
\end{array}\right),
\label{Hc}
\end{align}
where the index $l$ of the submatrix represents the spin of the light
degrees of freedom, whereas $h$ denotes that of the
heavy quark. The wave functions for the heavy field are then written
as 
\cite{Gupta:1993kd} 
\begin{align}
\bar{H}_{lh}^{b}=\left\{ \begin{array}{ll}
\frac{1}{\sqrt{4\pi M}}(\hat{n}\cdot\vec{\tau})_{bm}
                           \epsilon_{lm}u(r)\chi_{h} &
 \mbox{if }b=1,2,\\  \qquad\qquad0 & \mbox{if }b=3,
\end{array}\right.
\label{Hbar}
\end{align}
where $\chi_h$ stands for the heavy-quark spinor and $u(r)$ is the
radial wave function derived in the classical approximation, so that
the heavy quark is localized at the origin. Thus, it is given as 
$r^{2}|u(r)|^{2}\approx\delta(r)$, which is an exact expression in the
infite mass limit of the heavy quark ($m_Q\rightarrow\infty$). 

Substituting the Ansatz~(\ref{Hbar}) into the heavy-meson part of
the Lagrangian, we obtain the classical binding potential in nuclear
matter 
\begin{align}
W_{0}^*&=-\frac{3d}{2}F'(0)+\frac{3c}{m_v^*g^*}G''(0) - \frac{\alpha
         g^*}{\sqrt{2}}\omega(0). 
\label{BE}
\end{align}
For more details we refer to Ref.\,\cite{Gupta:1993kd}.

\subsection{Symmetry breaking terms}
In Ref.\,\cite{Schechter:1992iz}, it was shown in detail how the terms
for the explicit breaking of flavor SU(3) symmetry can be derived
based on the fundamental QCD Lagrangian\footnote{See Eq.(2.10) in
  Ref.\,\cite{Schechter:1992iz} and related discussions.}.  
The explicit breaking of chiral symmetry is also considered in
medium, since it plays an important role in justifying the medium 
modifications of the nonstrange baryons with regards to the
phenomenology of pion-nucleus scattering and the properties of pionic
atoms\,\cite{Ericsonbook,Rakhimov:1996vq} (for more discussions,  
see also Refs.\,\cite{Meissner:2006id,Yakhshiev:2013eya}). We assume
isospin symmetry. In addition, we also need to take into account the
explicit breaking of flavor SU(3) symmetry breaking. The corresponding
medium-modifed term in the effective Lagrangian is written
as\footnote{A free-space form of the Lagrangian is given in
  Ref.\,\cite{Gupta:1993kd}.}  
\begin{align}
\mathcal{L}_{SB}^{*}& =\frac{1}{8}\,m_\pi^{*2}F_\pi^{*2}
                      \mathrm{Tr}[\mathcal{M}(U+U^{\dagger}-2)]\cr 
&+\frac{M_s-M}{2(x-1)} [\Tr(H\xi{\cal M}\xi\bar{H})
+{\rm H.c.}],
\label{SB}
\end{align}
where $3\times 3$ matrix ${\cal M}$ with isospin symmetry
is given as 
\begin{align}
\mathcal{M}=T+xS
\end{align}
 with the diagonal matrices $T={\rm diag}(1,1,0)$ and 
 $S={\rm diag}(0,0,1)$.  The constant $x$ is defined by the raio of
 the currents quark masses 
\begin{align}
x=\frac{2m_{s}}{m_u+m_d}=31.5
\end{align}
and $M_s-M=m(D_s^+)-M(D^+)=100$\,MeV, which are accurate enough up to
2~\%\,\cite{Gupta:1993kd}. In the SU(2) sector the
Lagrangian~(\ref{SB}) reproduces the standard medium-modified chiral 
symmetry breaking term.  

\subsection{Zero-mode quantization}

The quatized solitons appear after the zero-mode quantization, which
is performed by rotating the classical static configuration in the
following way 
\begin{align}
\xi(\mathbf{x},t)&=A(t)\xi(\mathbf{x})A^{\dagger}(t),\cr
\rho_{\mu}(\mathbf{x},t)&=A(t)\rho_{\mu}(\mathbf{x})A^{\dagger}(t),\cr
\bar{H}(\mathbf{x},t)&=A(t)\bar{H}(\mathbf{x}),
\label{collective ansatzs}
\end{align}
where the classical Ans\"{a}tze are defined in Eqs.\,\eqref{xi},
\eqref{vecc} and \eqref{Hc}, and $A(t)$ denotes an SU(3) rotational
matrix in flavor space. The corresponding angular velocities
$\Omega_{K}$ of the rotation are defined by 
\begin{align}
A^{\dagger}\dot{A}=\frac{i}{2}\sum_{k=1}^{8}\lambda_{k}\Omega_{k},
\end{align}
where the $\lambda_{k}$ are the usual Gell-Mann matrices. 
Substituting Eq.~(\ref{collective ansatzs}) respectively into the
light \,(\ref{Llight}) and heavy \,(\ref{Lheavy}) parts of the
Lagrangian, we get the collective Lagrangian 
\begin{align}
L^*&=-M_{\mathrm{cl}}^* -
     (M^*+W^*_{0})P+\frac{\alpha^{2}}{2}\sum_{i=1}^{3}\Omega_{i}^{2}\cr  
&+\frac{\beta^{2}}{2}\sum_{k=4}^{7}\Omega_{k}^{2}-
  \frac{\sqrt{3}}{2}\Omega_{8}+\frac{\sqrt{3}}{6}\Omega_{8}
  \chi^\dagger\chi   P,   
\label{L*}
\end{align}
where the heavy meson mass in nuclear medium $M^*$ by added as an
overall energy shift. The medium-modified  classical soliton mass is
given by the expression\footnote{Note that  in the expression of the
classical soliton mass only the spatial parts of the constants 
$F_{\pi,s}^*$ and $g^*_s$ appear.}
\begin{align}
  M_{\rm cl}^{*}&=4\pi\int_0^\infty {\rm d}r\; \Bigg
                  \{\frac{F_{\pi,s}^{*2}}{4}(2\sin^{2}{F}+r^{2}F'^{2})\cr 
  &+m_{\pi}^{*2}F_{\pi,s}^{*2}r^{2}\sin^2\frac{F}{2}+\frac{k}{2}
    F_{\pi}^{2}(G-1+\cos{F})^{2}\cr 
  &+\frac{1}{2g^{*2}_s}\left(
    G'^{2}+\frac{G^{2}(G-2)^{2}}{2r^{2}}\right)\cr 
  &-\frac{r^{2}}{2}(m_{v}^{*2}\omega^{2}+\omega'^{2}) +
    \frac{3g^*_s}{2\sqrt{2}\pi^{2}}\omega F'\sin^{2}{F} \Bigg\}. 
  \label{Mcl*}
  \end{align}
Note that in formulating the classical soliton mass we have followed
the conventional way of including the symmetry breaking  
term (the pion mass term in $M_{\rm cl}^*$) in Eq.\,\eqref{SB}.
The classical equation of motion, its asymptotic solutions 
corresponding to the classical soliton configuration, the numerical
method, and the solutions are presented in Appendix\,\ref{app:EOM}.
The in-medium modified moments of inertia of the rotating soliton are
given by\footnote{Note that in the expressions of the moments of
  inertia only the temporal parts of the constants $F_{\pi,t}^*$ and
  $g^*_t$ appear.} 
\begin{align}
\label{alpha*}
  \alpha^{*2}&=\frac{4\pi}{3}\int {\rm d}r \Bigg\{F_{\pi,t}^{*2}
               r^{2}\sin^{2}{F}+\frac{2G^{2}}{g_t^{*2}}\cr 
  &+2kF_{\pi}^{2} r^{2}\sin^{4}{\frac{F}{2}}\Bigg\},\\
    \beta^{*2}&=2\pi\int {\rm d}r\Bigg\{F_{\pi,t}^{*2}
                r^{2}\sin^{2}{\frac{F}{2}}+\frac{G^{2}}{2g_t^{*2}}\cr 
    &+4kF_{\pi}^{2}r^{2}\sin^{4}{\frac{F}{4}}\Bigg\}.
    \label{beta*}
\end{align}
The factor $P$ in Eq.~\eqref{L*} is a projection operator onto the
heavy baryon subspace. When $P$ picks only up the heavy-quark sector.
$W_{0}$ is the classical binding energy given in
Eq.\,\eqref{BE}. The last term in Eq.\,\eqref{L*} originates from 
the kinetic term for the heavy mesons,
i.e. $iV_{\mu}\mathrm{Tr}(H\partial_{\mu}\bar{H})$.  
The second lower index $h$ in the heavy super-field~\eqref{Hbar}  
stands for the spin of the heavy-quark degree of freedom, so that the
heavy-quark spinor satisfies the normalization
$\chi^{\dagger}\chi=\delta_{hh'}$ that appears as the  
last term in Eq.\,\eqref{L*}. Since the heavy-quark spin is conserved
in the infinite mass limit of the heavy quark, the term with
$\chi^\dagger \bm{\sigma} \chi$ vanishes. 
This indicates that the heavy quark spin is considered to be 
a good quantum number for the collective Lagrangian. Thus, the
collective wave functions, which will be discussed later, retain the
heavy-quark spinor. When $m_Q\to \infty$, the heavy quark does not
contribute to any dynamical properties of the singly heavy baryons but
is only regarded as a static color source.   

The collective Lagrangian does not have the quadratic terms
for $\Omega_8$, since both the chiral field $U(\mathbf{r})$ and the
heavy super-field commute with $\lambda_{8}$. The linear-order term
with $\Omega_8$ is associated with the baryon number, which we now
discuss. Introducing the canonical conjugate momenta $R_{k}$
\begin{align}
-R_{k}=\frac{\partial L}{\partial\Omega_{k}},\qquad k=1,\cdots 7,
\end{align}
we find 
\begin{align}
R_{k}=\left\{ \begin{array}{ll}
-\alpha^{2}\Omega_{k}, & \;\;\; k=1,2,3\\
-\beta^{2}\Omega_{k}, & \;\;\; k=4,5,6,7\\
\quad\frac{1}{\sqrt{3}}, & \;\;\; k=8.
\label{qcondition}
\end{array}\right.
\end{align}
The $8^{\rm th}$ component of $R_k$ is constrained by the WZW
term and the heavy meson kinetic term. In the usual SU(3) Skyrme model  
for light baryons, the baryon number arises from the WZW term. So,
$R_8$ is constrained to be $R_{8}=\frac{\sqrt{3}}{2}$, which is
identified as the right hypercharge $Y_R=\frac{2}{\sqrt{3}}
R_8=N_c/3$. This constraint allows one to take only the SU(3)
representation with zero triality. When we consider the singly heavy
baryons as in this work, however, the effect of the kinetic term in
the heavy Lagrangian~\eqref{Lheavy} constrains further
$R_{8}=\frac{1}{\sqrt{3}}$. This has a profound physical meaning. 
The right hypercharge for the singly heavy baryons is then
$Y_R=(N_c-1)/3$ and this allows us to take the presentations for the
singly heavy baryons $\bar{\bm{3}}$ with $J=1/2$ and $\bm{6}$ with
$J=1/2$ and $J=3/2$. Moreover, the soliton for the singly heavy
baryons appears as that consisting of $N_c-1$. Thus, even  
though the present work does not contain the explicit quark degrees of 
freedom, it has a certain similarity to the pion mean-field
approach~\cite{Yang:2016qdz} developed recently.

We then arrive at the collective Hamiltonian of the model in the
heavy sector 
\begin{align}
  H&=M_{\rm cl}^*+(M^*+W_{0}^*)P\cr
     &+\frac{1}{2}\bigg(\frac{1}{\alpha^{*2}} -
   \frac{1}{\beta^{*2}}\bigg)J_{s}(J_{s}+1)\cr  
  &+\frac{1}{6\beta^{*2}}\left[(p^{2}+pq+q^{2})+3 (p+q)\right]\cr
  &-\frac{1}{2\beta^{*2}}R_{8}^{2} 
\label{Hamil} 
\end{align}
in the $(p,q)$ representation of flavor SU(3) symmetry. Diagonalizing
the collective Hamiltonian, we find the collective wave functions for
the singly heavy baryons as~\cite{Momen:1993ax, Yang:2018uoj}
\begin{align}
\Psi_{\rm{heavy}}({n},YII_{3};JJ_{3},J_{s};A) =
  \sum_{h=1}^{2}C^{J_{s}\ \frac{1}{2}\ J}_{M_{s}\ h\ J_{3}}\cr 
\times(-1)^{J_{s}-M_{s}}\sqrt{\mathrm{dim}\ n}
  D^{(n)*}_{Y~I~I_{3},\ Y_{R}~J_{s}~-M_{s}}(A)\chi_{h}, 
\label{wavefunction}
\end{align}
where $J_{s}$ and $M_{s}$ are the soliton spin and its $3^{rd}$
component. $C^{J_{s}\ \frac{1}{2}\ J}_{M_{s}\ h\ J_{3}}$ denote the
SU(2) clebsch-Gordan coefficients that couple the soliton and the
heavy quark. 

The effects of explicit flavor SU(3) symmetry breaking are taken as
a perturbation~\cite{Gupta:1993kd}. Then the collective Hamiltonian
can be written as 
\begin{align}
H = H_{\mathrm{sym}} + H_{\mathrm{br}},
\end{align}
where the part from explicit SU(3) symmetry breaking is expressed as
\begin{align}
H_{\mathrm{br}} &=\tau D_{88}(A),\qquad \tau =\tau_{\rm
                  light}^*+\tau_{\rm heavy}^*, \\ 
\tau^{*}_{\rm light}&=\frac{2\pi}{3}(1-x)m_{\pi}^{*2}F_{\pi}^{*2}
\int_0^\infty {\rm d}r\;r^{2}\sin\frac{F}{2}, \\
 \tau_{\rm heavy}^*&=
\frac{M_s^*-M^*}{3}.
\label{eq:exbr}
\end{align}
Here $M_s$ means the heavy meson mass with the $s$ quark.  
The value of $\tau_{\rm heavy}$ in free space is fixed by the relation
$M_s-M=m(D_s^{+*})-m(D^{+*})=m(D_s^{+})-m(D^{+})=100$\,MeV\,\cite{Gupta:1993kd}.   

\section{Nuclear matter}
For simplicity we will consider isospin symmetric infinite nuclear 
matter in order to discuss the medium effects on the baryon
properties. We follow the method presented already  
in Refs.\,\cite{Jung:2012sy,Yakhshiev:2013eya} and introduce the 
medium functions as 
\begin{align}
F_{\pi,t}^{*}=\sqrt{\alpha_{p,t}}F_{\pi},\quad g^{*}_t=\sqrt{\zeta_t}g,\cr
F_{\pi,s}^{*}=\sqrt{\alpha_{p,s}}F_{\pi},\quad g^{*}_s=\sqrt{\zeta_s}g.
\end{align}
The masses of the pseudoscalar and vector mesons undergo the changes
in nuclear matter, so that we consider the medium effects for them
as~\cite{Jung:2012sy,Yakhshiev:2013eya} 
\begin{align}
m_\pi^*=\sqrt{\frac{\alpha_m}{\alpha_{p,s}}}m_\pi,\qquad
  m_{v}^{*}=\sqrt{\zeta_s}m_{v}. 
\end{align}
In addition, we define four different medium functions by three
constants in the following way~\cite{Yakhshiev:2013eya} 
\begin{align}
\alpha_{p,s} &=1+C_{1}\lambda\,\quad
\zeta_{s} = 1+C_{2}\lambda\,,\cr
\alpha_{p,t} &= 1+C_{3}\lambda\,\quad
\zeta_{t} = \alpha_{p,s}\zeta_{s}\alpha_{p,t}^{-1}.
\label{param}
\end{align}
The remaining medium functions will not affect much nuclear matter
properties, so that they are chosen to be~\cite{Jung:2012sy}  
\begin{align}
\alpha_m=1-\frac{4\pi b_0\eta\rho}{m_\pi^2},\,\quad
  \eta=1+\frac{m_\pi}{m_N}, 
\end{align}
where $b_0=-0.024m_\pi^{-1}$ is taken from the data on pionic
atoms~\cite{Ericsonbook}.   

We fit the values of the constants $C_{1,2,3}$ by reproducing the bulk
properties of isospin symmetric nuclear matter near the normal
nuclear-matter density $\rho_0=0.16\,{\rm fm}^{-3}$. 
We introduce the normalized nuclear matter density 
$\lambda=(\rho_{\rm p}+\rho_{\rm n})/\rho_0$ in terms of the 
proton and neutron distribution densities.  Then the volume term 
in the binding energy formula is defined by 
\begin{align}
\varepsilon_V(\lambda)&=\frac{ZM_{p}^{*}(\lambda)
 +NM_{n}^{*}(\lambda)}{A}-\frac{ZM_{p}+NM_{n}}{A}\cr  
&=M_{N}^*(\lambda)-M_N, 
\end{align} 
where 
\begin{align}
M_{N}^{*}&=\frac{M_p^*(\lambda)+M_n^*(\lambda)}2\cr
&=M_{\rm cl}^{*}(\lambda)+\frac{3}{8}\frac{1}{\alpha^{*2}(\lambda)} +
  \frac{3}{4}\frac{1}{\beta^{*2}(\lambda)}\cr 
&-\tau_{\rm light}^{*}(\lambda)\left(1-\frac{3}{10}\right).
\end{align}
Utilizing the stability condition $(\partial
\varepsilon_V(\lambda)/\partial\lambda)_{\lambda=1}=0$, we find the volume
energy value $\varepsilon_V(1)= 16$\,MeV and the compressibility of
the symmetric matter $K_0=\{9\lambda^{2}({\partial^{2}\varepsilon_V}/ 
{\partial\lambda^{2}})\}_{\lambda=1}=220$\,MeV, which are used for  
fixing the values of constants in the density  
functions in Eq.\,(\ref{param})
\begin{align}
C_1&=-0.130275,\quad\cr
C_2&=~\,\,0.488595,\quad\\
C_3&=-0.203271.\nonumber
\end{align}

\begin{figure}[htp]
\includegraphics[scale=0.19]{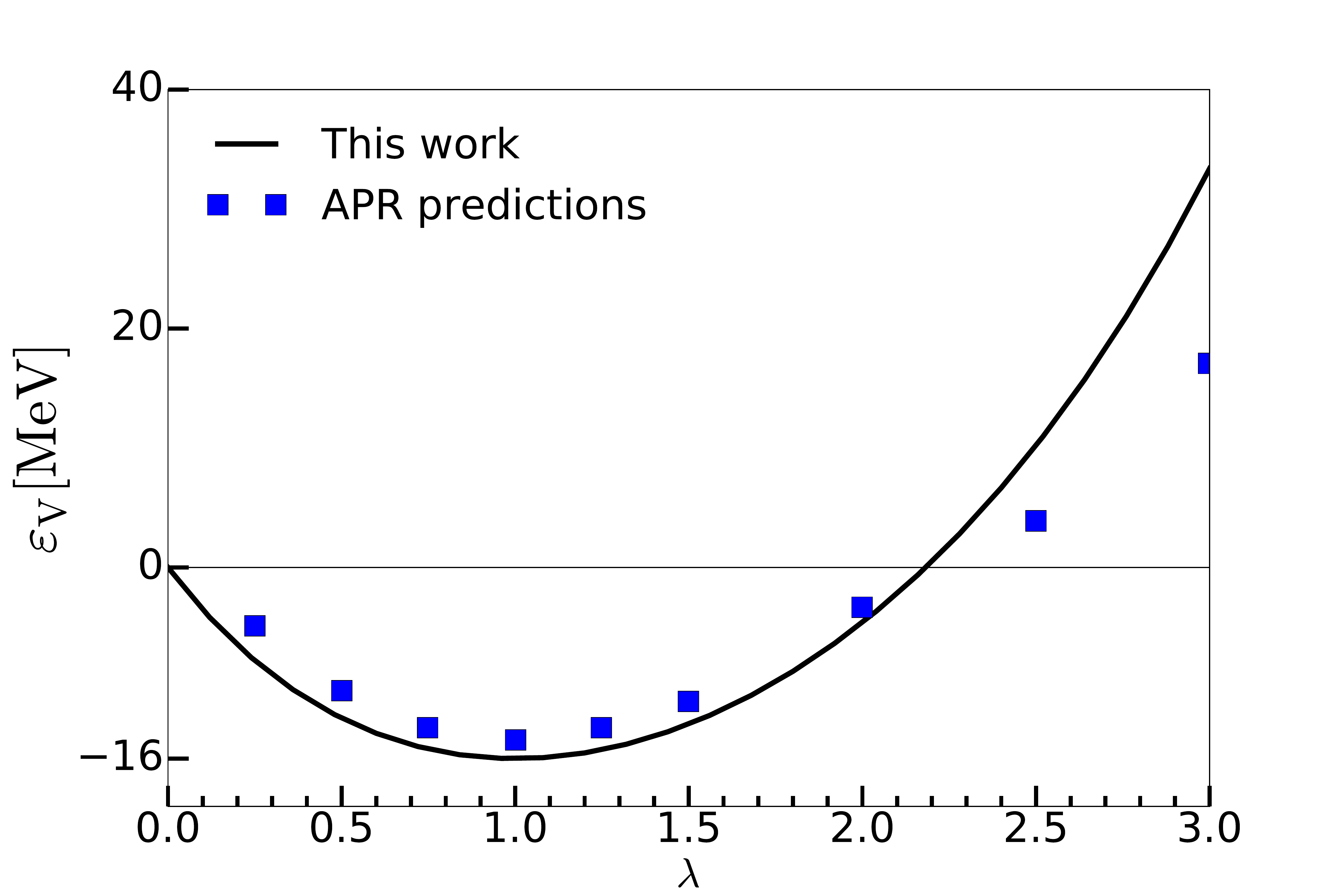}
\caption{The dependence of the volume energy on the normalized 
nuclear matter density $\lambda=\rho/\rho_0$.
The solid curve depicts the result from this work. The squares  
represent the Akmal-Pandharipande-Ravenhall (APR)
predictions\,\cite{Akmal:1998cf}.}
\label{fig1}
\end{figure}

The masses of baryons in this model are quite
overestimated, e.g. the nucleon  mass in free space is  $M_N=
2094$\,MeV. This is well known as a fundamental problem in any chiral
soliton models. Thus, instead of using this value, we fix the model
parameters in free space such that the masses of the singly heavy
baryons are reproduced properly in free space. See Table~\ref{tab:1}
for the values of singly heavy-baryon masses in free space in
comparison with the experimental data. Following
Ref.\,\cite{Jung:2012sy}, we introduce the scaling factor 
$f_s=M_N/M_N^{\rm exp} =2094/940=2.228$ and use it
to reproduce nuclear matter properties at the normal nuclear matter
density $\rho_0$. The density dependencies of the volume energy  
for isospin symmetric matter is shown in Fig.~\ref{fig1}.  
The solid curve draws the present results including the inverse scale
factor $f_s^{-1}$, where the binding energy per nucleon at the normal
nuclear matter density is given by $a_0=\varepsilon_V(1)=-16\,$MeV. 
It is compared with the Akmal-Pandharipande-Ravenhall (APR) 
predictions\,\cite{Akmal:1998cf} denoted by the blue squares.  One can
see that the present results are in good agreement with the APR
predictions in the density region $\rho\in[0,2\rho_0]$.  
When the value of the compressibility is taken to be $K_0=220$\,MeV, 
we get the third derivative parameter of equations of state
$Q_0=-309.927$\,MeV, which is in qualitative agreement with many other
approaches.  

\begin{figure}[htp]
\includegraphics[scale=0.19]{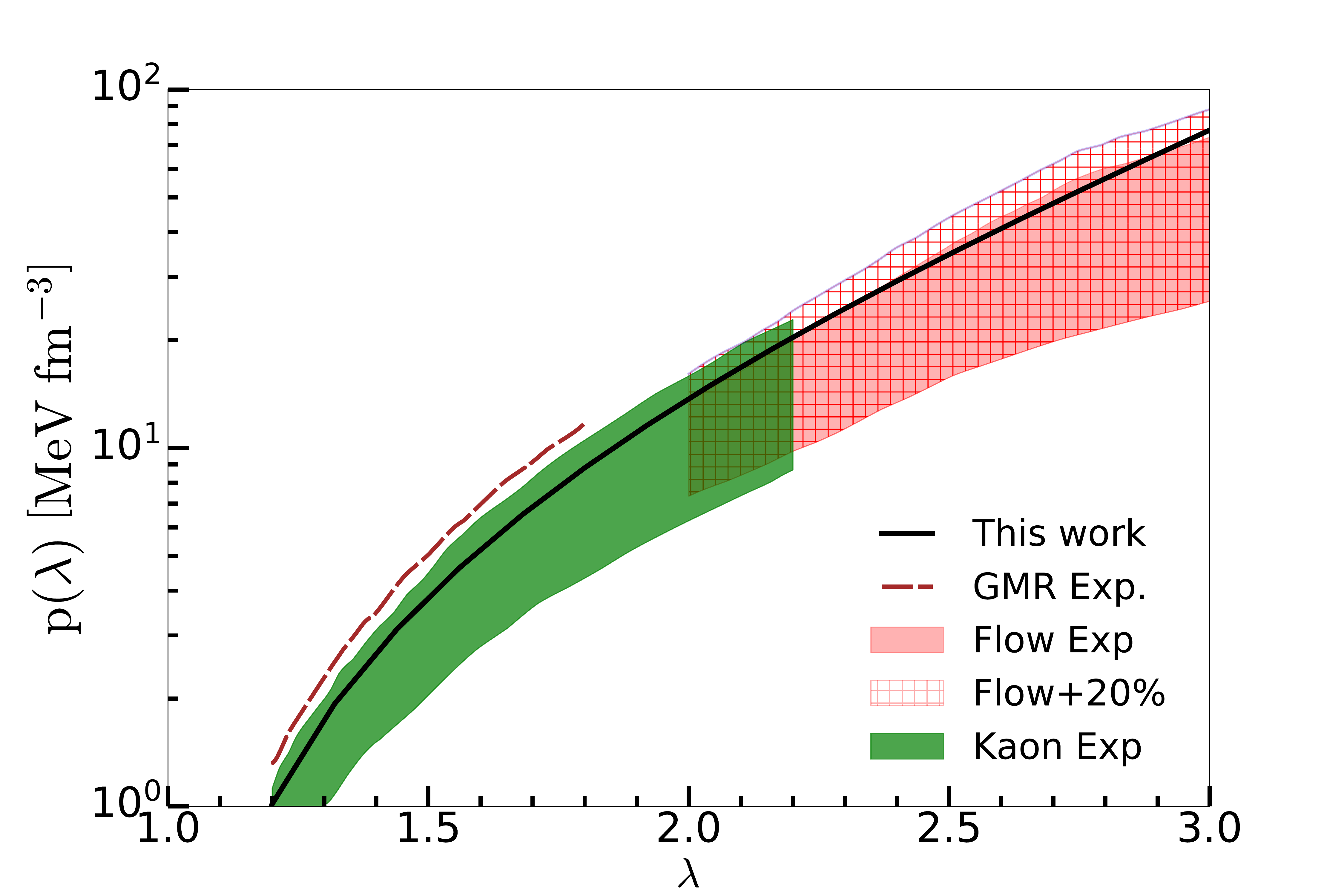}
\caption{The dependence of the pressure on 
the normalized nuclear matter density
$\lambda=\rho/\rho_0$. The solid cuve depicts the present results. The 
data are taken from GMR\,\citep{Youngblood:1999zza,Lynch:2009vc}, 
flow\,\citep{Danielewicz:2002pu}, flow$+20
\%$\,\citep{Steiner:2012xt,Dutra:2014qga}, and
kaon\,\citep{Fuchs:2005zg,Lynch:2009vc} experiments.
}
\label{fig2}
\end{figure}

The density dependence of the pressure in isospin symmetric
matter is shown 
in Fig.~\ref{fig2}.  
While the pressure in isospin symmetric matter 
 is proprtional to the first derivative of the volume energy  
$p(\lambda)=\rho_0\lambda^2(\partial\varepsilon/
\partial\lambda)$, the same rule (inverse scale factor
$f_s^{-1}$) also holds for it. The present result for the pressure 
is found to be within the allowed region, given the range of the
density $\rho\in[0,3\rho_0]$. In Fig.\,\ref{fig2}, 
we draw also the data extracted from various 
experiments. The present result is slightly lower 
than the data from the giant monopole resonance (GMR)
experiments\,\citep{Youngblood:1999zza,Lynch:2009vc}    
for heavy nuclei, which are depicted by the long-dashed curve. 
Reference\,\citep{Danielewicz:2002pu} analyzed 
the flow experimental data on ${}^{197}\mathrm{Au}$ nuclei
collision, which is illustrated by the red-colored region
corresponding to the equation of state for symmetric nuclear matter at
zero temperature. The valid range of the pressure was extended  
by taking into account the mass-radius relation of neutron stars from
observational data~\citep{Steiner:2012xt,Dutra:2014qga}. 
In Fig.~\ref{fig2}, the corresponding region is denoted by
"$\mathrm{Flow}+20\%$". One can see that the present equation of
state lies within the predicted range. 
The data taken from the kaon production at high-energy
nucleus-nucleus collision~\cite{Fuchs:2005zg,Lynch:2009vc}
are drawn as the green-colored region in the range of  $1.2 \leq
\lambda \leq 2.2$.  The present result is also located within this
region. We have also considered the two different values of
$K_0$, i.e., $K_0=240\,$MeV and $K_0=260\,$MeV but the results for the
pressure are also in agreement with these data. Thus, the present
theoretical framework describes the bulk properties of symmetric
nuclear matter rather well. 

\section{Singly heavy baryons in nuclear medium}
We now proceed to investigate the singly heavy baryons in nuclear
medium. Since the mass of the heavy meson $M^*$ is involved in the
Hamiltonian Eq.~\eqref{Hamil} as an overall energy shifting factor, 
we have to examine how the masses of the heavy
mesons undergo the change, i.e., to see how $\Delta M^*=
M^*-M$ depends on the nuclear density.  
In the present work $M$ is taken to be the $D$-meson mass $M_D$.
The properties of heavy mesons and in particular those of the $D$
meson in nuclear environment are discussed within the several 
theoretical approaches\,\cite{Lutz:2005vx, Chhabra:2017rxz,
  Chhabra:2017emy, Krein:2017usp, Suenaga:2017deu, Suzuki:2015est,
  Pathak:2014vra,Yasui:2012rw,Hilger:2008jg}.  A general consensus has
not been reached yet, whether the $D$-meson mass  
would be dropped or raised in nuclear matter. Thus, we will consider
in this work three possible cases of the changes of the $D$ meson in
the following. 
We introduce one more density-dependent constant $C_4$ to parametrize
the density dependence of the $D$-meson mass in the following linear
form 
\begin{align}
M^*\equiv M_{D}^{*}=(1+C_{4}\lambda)M_{D}.
\label{MD*}
\end{align}
Moreover, since we focus on the properties of isospin symmetric
matter, we ignore the isospin effects on the masses of the heavy mesons.
This means that the density dependence of the symmetry breaking
coefficient from the heavy-quark sector $\tau_{\rm heavy}^*$ given in
Eq.~\eqref{eq:exbr} should also be modified:
\begin{align}
\tau_{\rm heavy}^*=(1+C_{4}\lambda)\tau_{\rm heavy}.
\label{tauH*}
\end{align}
In the present work we consider $C_4$ as a free parameter and examine
how the masses of the singly heavy baryons vary with $C_4$ changed. As
mentioned previously, We will take the three different cases: the
increment of the $D$-meson mass ($C_4>0$), no change
($C_4=0$), and the dropping mass of the $D$ meson ($C_4<0$) in nuclear
matter. 

We are now in a position to discuss the mass shifts of the
singly heavy baryons $\Delta M_B= M_B^*-M_B$ in nuclear matter. We
focus our attention on the properties of the singly charmed
baryons. Given the different values of free parameter $C_4$, 
the changes of the heavy baryon masses in nuclear matter at the normal
nuclear matter 
density are shown in Table~\ref{tab:1}. 
\begin{table}[hbt]
\caption{The masses of the singly heavy baryons in free space and
  their shifts in nuclear matter $\Delta M_B$ at the normal nuclear
  matter density, $\rho=\rho_0$. The experimental values of the masses in
  free space are also given. All the masses and their shifts are given
  in units of MeV.} 
\begin{center}
\renewcommand{\arraystretch}{1.3}
\scalebox{1.0}{%
\begin{tabular}{c|cc|ccc}
\hline
\hline
&\multicolumn{2}{c|}{$M_B$, $\rho=0$ }&
\multicolumn{3}{c}{ $\Delta M_B$, $\rho=\rho_0$}\\
\cline{2-6}
Baryon & Exp.\cite{Agashe:2014kda}  & This work
& $C_4=-0.1$ & $C_4=0$ & $C_4=0.1$ 
  \\
\hline
${\Lambda_{c}}$ & 2286.5 & 2286.0 & $-166.91$ & 21.22   & 209.34\\
${\Xi_{c}}$   & 2469.4   & 2437.8 & $-132.30$ & 54.48   & 241.25 \\
\hline
${\Sigma_{c}}$ & 2453.5  & 2564.5 
& $-86.50$ & 101.54 & 289.57 \\
${\Xi_{c}^{\prime}}$ & 2576.8 & 2646.8 
& $-69.89$  & 117.27 & 304.43  \\
${\Omega_{c}}$ & 2695.2  & 2721.6 
& $-54.23$  & 132.17 & 318.56 \\
\hline
\hline
\end{tabular}}
\end{center}
\label{tab:1}
\end{table}
We also provide the values of the masses of the singly heavy baryons
in free space and the corresponding experimental
data~\cite{Agashe:2014kda}.  As mentioned above, the parameters of the
model in free space are fitted to the experimental data on the masses
of the baryons. One can see that the tendency in the change of the
baryon masses turns out in different ways, depending on the value of free
parameter $C_4$.  For example, if one chooses $C_4=-0.1$ 
that corresponds to the dropped mass of the $D$ meson in nuclear matter,
then the masses of the singly heavy baryons also fall off linearly in
nuclear matter. If the mass of the $D$ meson does not change in
nuclear matter $C_4=0$, then the baryon masses slightly increase in
nuclear matter. With the positive values of $C_4$ taken, the masses of
the singly heavy baryons rises linearly.

\begin{figure}[htp]
\includegraphics[scale=0.19]{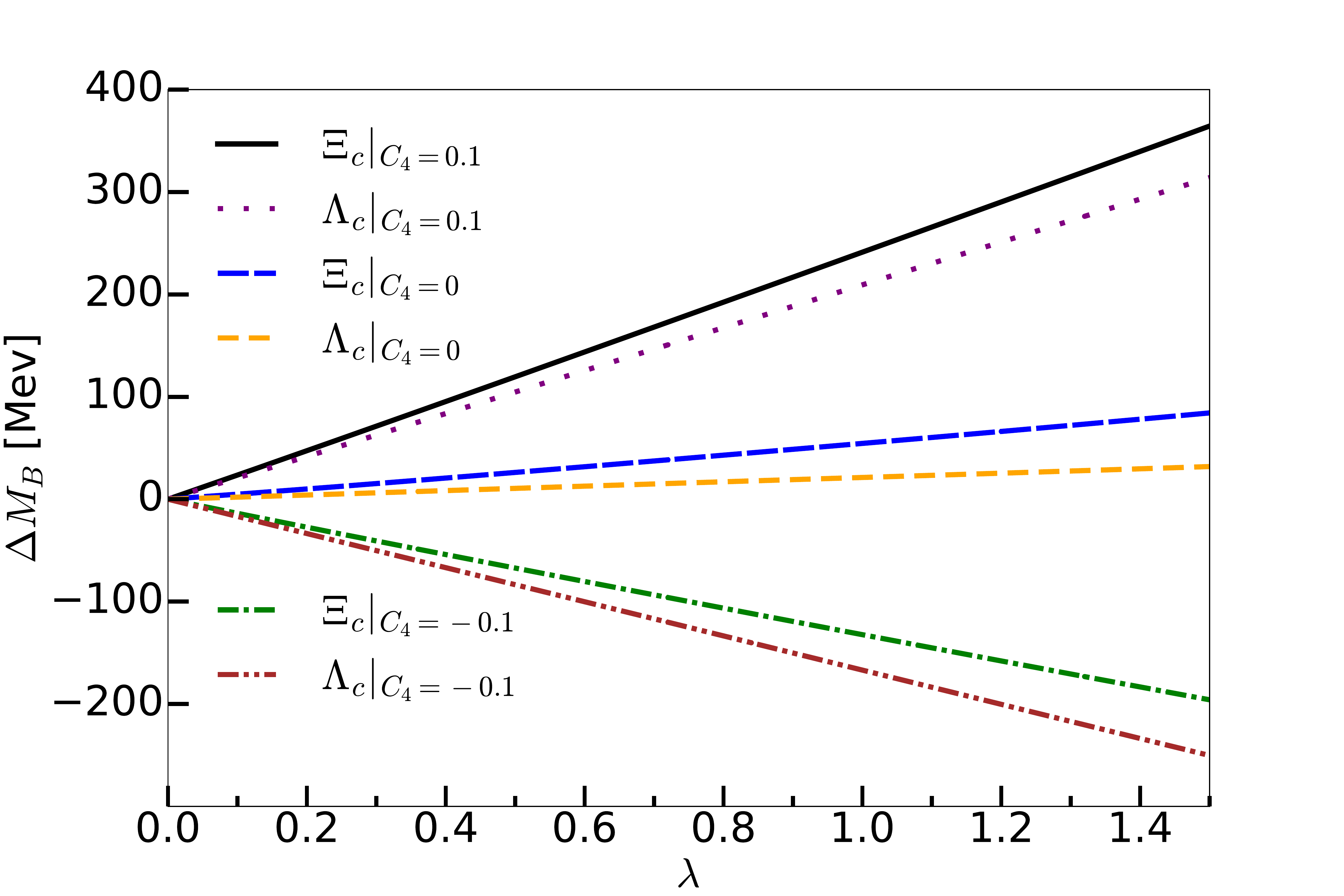}
\caption{Mass shifts for the baryon antitriplet in nuclear matter
  $\Delta M_B$ as functions of the reduced nuclear matter density
  $\lambda$ with three different values of $C_4$ taken, i.e.,
  $C_4=0.1$, $C_4=0$ and $C_4=-0.1$. }
\label{fig3}
\end{figure}

The density dependence of the mass shifts for the baryon antitriple
$\Delta M_B$ are shown in Fig.~\ref{fig3}. 
One can see that the masses of the baryon antitriplet are linearly
proportional to the values of $C_4$. 
We find that those of the baryon sextet exhibit a similar tendency,
which are drawn in Fig.~\ref{fig4} with the two different values of
$C_4$ given. 
\begin{figure}[htp]
  \includegraphics[scale=0.19]{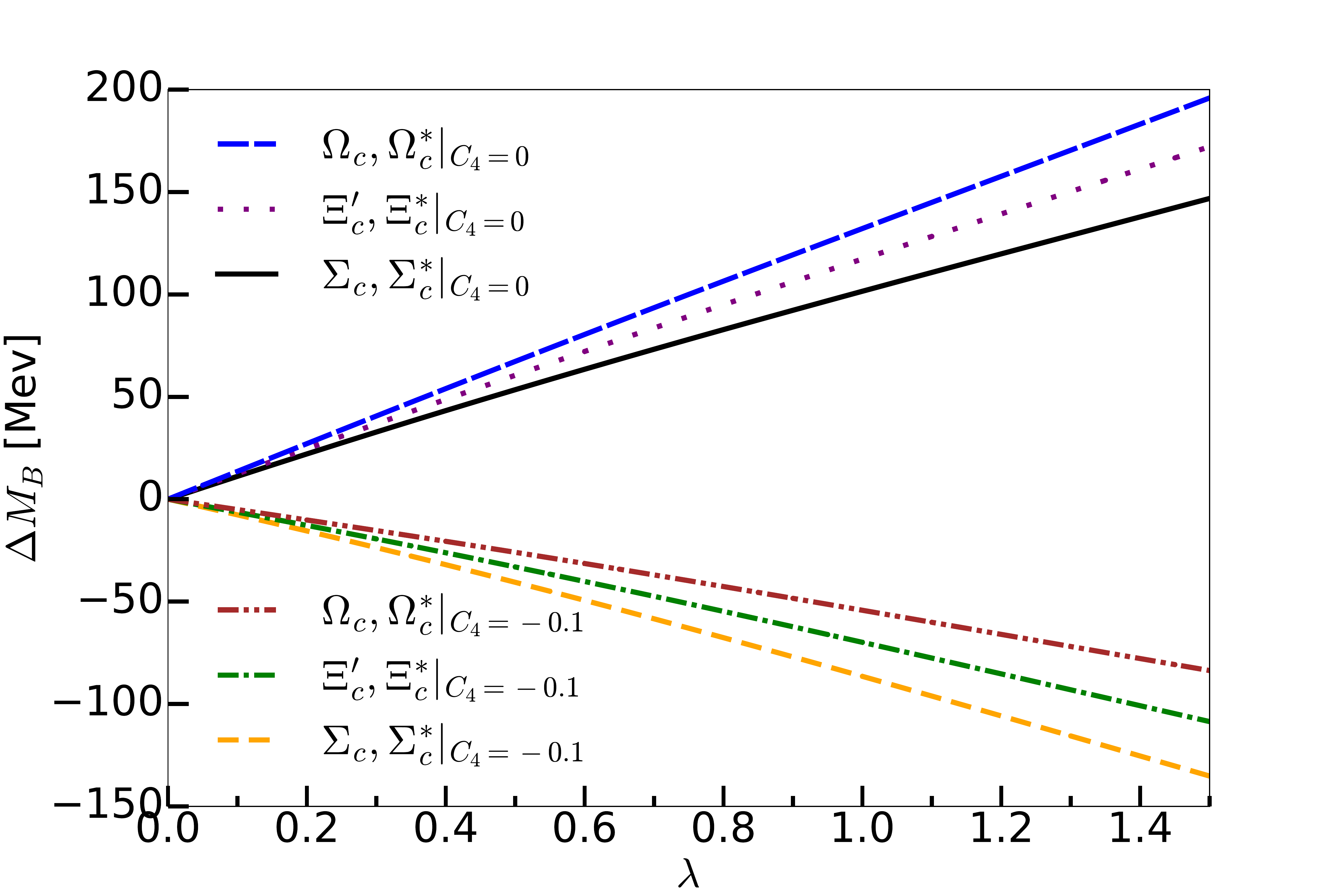}
\caption{Mass shifts for the baryon sextet in nuclear matter
  $\Delta M_B$ as functions of the reduced nuclear matter density
  $\lambda$ with two different values of $C_4$ taken, i.e.,
  $C_4=0$ and $C_4=-0.1$.}
\label{fig4}
\end{figure}
The masses of the baryon sextet drop less than those of the baryon
antitriplet with the negative values of $C_4=-0.1$ taken. In the case
of the null or positive values of $C_4$, those of the baryon sextet
grow slower in nuclear matter than those of the baryon antitriplet.

In order to analyze the mass splittings between the baryon sextet with
spin 1/2 and 3/2, we need to introduce the hyperfine interaction with
the anomalous chromomagnetic moment~\cite{Yang:2016qdz}
\begin{align}
H_{LH}^{*}&=\frac{2}{3}\frac{\kappa}{m_{c}M_{\rm cl}^{*}}
            \mathbf{S}_{L}\cdot\mathbf{S}_{H} 
\equiv\frac{2}{3}\frac{\xi}{m_{c}}\frac{M_{\rm cl}}{M_{\rm cl}^{*}}
            \mathbf{S}_{L}\cdot\mathbf{S}_{H}, 
\label{eq:hyp}
\end{align}
where $\kappa$ denotes the the anomalous chromomagnetic moment and
$m_c$ stands for the charm quark mass. We assume that $\kappa$ is not
influenced in nuclear matter. $\mathbf{S}_{L,H}$ designate the spins of
the soliton and heavy quark coming from the heavy meson, respectively.  
We introduce the coefficient $\xi$ with $M_{\rm cl}$ in free space.
We fit the value of $\xi/m_{c}=\kappa/(m_cM_{\rm cl})
\simeq 68.1$ MeV in free space to the experimental values of the
masses for the baryon sextet.

The mass shift of the baryon sextet including the hyperfine splitting
in nuclear matter are listed in Table~\ref{tab:2}.
\begin{table}[hbt]
\caption{The masses of the baryon sextet in free space and
  their shifts in nuclear matter $\Delta M_B$ at the normal nuclear
  matter density, $\rho=\rho_0$, with the hyperfine mass splitting
  considered. The experimental values of the masses in free space are
  also given. All the masses and their shifts are given in units of
  MeV.} 
\begin{center}
\renewcommand{\arraystretch}{1.3}
\scalebox{1.0}{%
\begin{tabular}{c|cc|ccc}
\hline
\hline
&\multicolumn{2}{c|}{$M_B$, $\rho=0$ }&
\multicolumn{3}{c}{ $\Delta M_B$, $\rho=\rho_0$}\\
\cline{2-6}
Baryon & Exp.\cite{Agashe:2014kda} & This work
& $C_4=-0.1$ & $C_4=0$ & $C_4=0.1$  \\
\hline
${\Sigma_{c}}$& 2453.5  & 2564.5 
& $-93.42$ & 94.62 & 282.65 \\
${\Xi_{c}^{\prime}}$ & 2576.8 & 2646.8 
& $-76.80$  & 110.36 & 297.51  \\
${\Omega_{c}}$ & 2695.2  & 2721.6 
& $-61.15$  & 125.25 & 311.64 \\
\hline
${\Sigma_{c}^{*}}$ & 2518.1  & 2587.2 
& $-83.04$ & 105.00 & 293.03 \\
${\Xi_{c}^{*}}$ & 2645.9 & 2669.6 
& $-66.43$  & 120.73 & 307.89  \\
${\Omega_{c}^{*}}$ & 2765.9  & 2744.3 
& $-50.77$  & 135.62 & 322.02 \\
\hline
\hline
\end{tabular}}
\end{center}
\label{tab:2}
\end{table}
One can see that if the anomalous chromomagnetic moment $\kappa$ and
the current quark mass $m_c$ are density independent, hyperfine mass
splittings are enhanced in nuclear matter because of the soliton mass
shown in Eq.~\eqref{eq:hyp}.  

\section{Summary and conclusions}
In the present work we investigated the medium modification of
heavy baryon masses in nuclear matter within the framework of the
SU(3) soliton model including the light pseudoscalar and vector
mesons, and the heavy meson. We introduced the medium effects in the
effective chiral Lagrangian and examined the bulk properties of
nuclear matter. The results are consistent with the empirical and
experimental data. It was shown that the results for the
mass shifts of the singly heavy baryons depend strongly on the medium
modification of the heavy meson mass in nuclear matter. In particular,
the heavy baryon masses gradually increase if the heavy meson
mass remains constant in nuclear matter. On the other hand, if the
heavy meson mass becomes larger in nuclear matter, then the baryon 
masses grow faster as the nuclear matter density increases.
With the mass of the heavy meson decreased in nuclear matter, we
found that the masses of the singly heavy baryons also lessen as the
nuclear matter density increases. We conclude that the mass dropping
of the singly heavy baryons crucially depends on information about how
the mass of the heavy meson undergoes changes in nuclear matter within
the present framework. 

\section*{Acknowledgments}
The present work was supported by Basic Science Research Program
through the National Research Foundation of Korea funded by the Korean
government (Ministry of Education, Science and Technology, MEST),
Grant-No. 2021R1A2C2093368, 2018R1A5A1025563 (H.-Ch.K.), and
2020R1F1A1067876 (U. Y.). 

\appendix
\section{Equations of Motion}
\label{app:EOM}
Equations for the profile functions $F$, $G$ and $\omega$ are obtained by 
the minimization of the classical soliton mass Eq.\,(\ref{Mcl*}) and
have the following forms 
\begin{align}
F^{\prime\prime}&=-\frac{2}{r}F^{\prime}-\frac{1}{\alpha_{p,s}r^{2}}[2k(G-1)\sin{F}\cr
&+(k-\alpha_{p,s})\sin{2F}]+\frac{\alpha_{m}}{\alpha_{p,s}}m_{\pi}^{2}\sin{F}\cr
&-\frac{\zeta_{s}^{1/2}}{\alpha_{p,s}}\frac{6g}{2\sqrt{2}\pi^{2}F_{\pi}}\frac{\bar{\omega}'}{r^{2}}\sin^{2}{F},\\
G^{\prime\prime}&=\frac{G\left(G-1\right)\left(G-2\right)}{r^{2}}\cr
&+\zeta_{s}m_{v}^{2}\left(G-1+\cos{F}\right),\\
\bar{\omega}^{\prime\prime}&=-\frac{2}{r}\bar{\omega}^{\prime}+\zeta_{s}m_{v}^{2}\bar{\omega}\cr
&-\zeta_{s}^{1/2}\frac{3g}{2\sqrt{2}\pi^{2}F_{\pi}}\frac{F^{\prime}}{r^{2}}\sin^{2}{F},  
\end{align}
where $\bar{\omega}$ is defined as $\bar{\omega}=\omega/F_{\pi}$.
The solution near the origin $r\rightarrow 0$ is found to be 
\begin{align}
F&=\pi+\alpha_{F}r,\quad
G=2+\alpha_{G}r^{2},\quad
\bar{\omega}=\bar{\omega}_{0}+\alpha_{\bar{\omega}}r^{2}.
\end{align}
At large distances $r \rightarrow\infty $ one gets 
\begin{align}
F&=\frac{\beta_{F}}{r^{2}}\left(1+\sqrt{\frac{\alpha_{m}}{\alpha_{p}^{s}}}m_{\pi}r\right)e^{-\sqrt{\frac{\alpha_{m}}{\alpha_{p}^{s}}}m_{\pi} r},\\
G&=\frac{\beta_{G}}{r^{2}}e^{-2\sqrt{\frac{\alpha_{m}}{\alpha_{p}^{s}}}m_{\pi} r},
\quad \bar{\omega}=\frac{\beta_{\bar{\omega}}}{r^{3}}e^{-3\sqrt{\frac{\alpha_{m}}{\alpha_{p}^{s}}}m_{\pi} r}.
\end{align}
The constants $\alpha_{F,G,\bar{\omega}}$ and $\beta_{F,G,\bar{\omega}}$ are 
found by the numerical calculations. 

\section{Solutions}

The numerical calculations for finding the values of $C_{1,2,3}$ are performed 
by an iteration method.  As an initial step the solutions in
free space are used (see the first row in Table~\ref{table3}).
\begin{table}[htp]
\caption{The values of medium parameters from the iterations.  }
\renewcommand{\arraystretch}{1.3}
\begin{tabular}{cccc}
\hline
\hline
Iteration number & $C_{1}$ & $C_{2}$ & $C_{3}$\\
\hline
0  \;&\; $-0.119060$ \;&\; 0.460658 \;&\; $-0.171721$  \\
3  \;&\; $-0.130249$ \;&\; 0.488556 \;&\; $-0.202932$ \\
8  \;&\; $-0.130275$ \;&\; 0.488595 \;&\; $-0.203271$  \\
9  \;&\; $-0.130275$ \;&\; 0.488595 \;&\; $-0.203271$ \\
\hline\hline
\end{tabular}
\label{table3}
\end{table}
The corresponding solutions in free space and in nuclear matter at the
saturation density are shown in Fig.~\ref{fig4}.
\begin{figure}[htp]
  \includegraphics[scale=0.19]{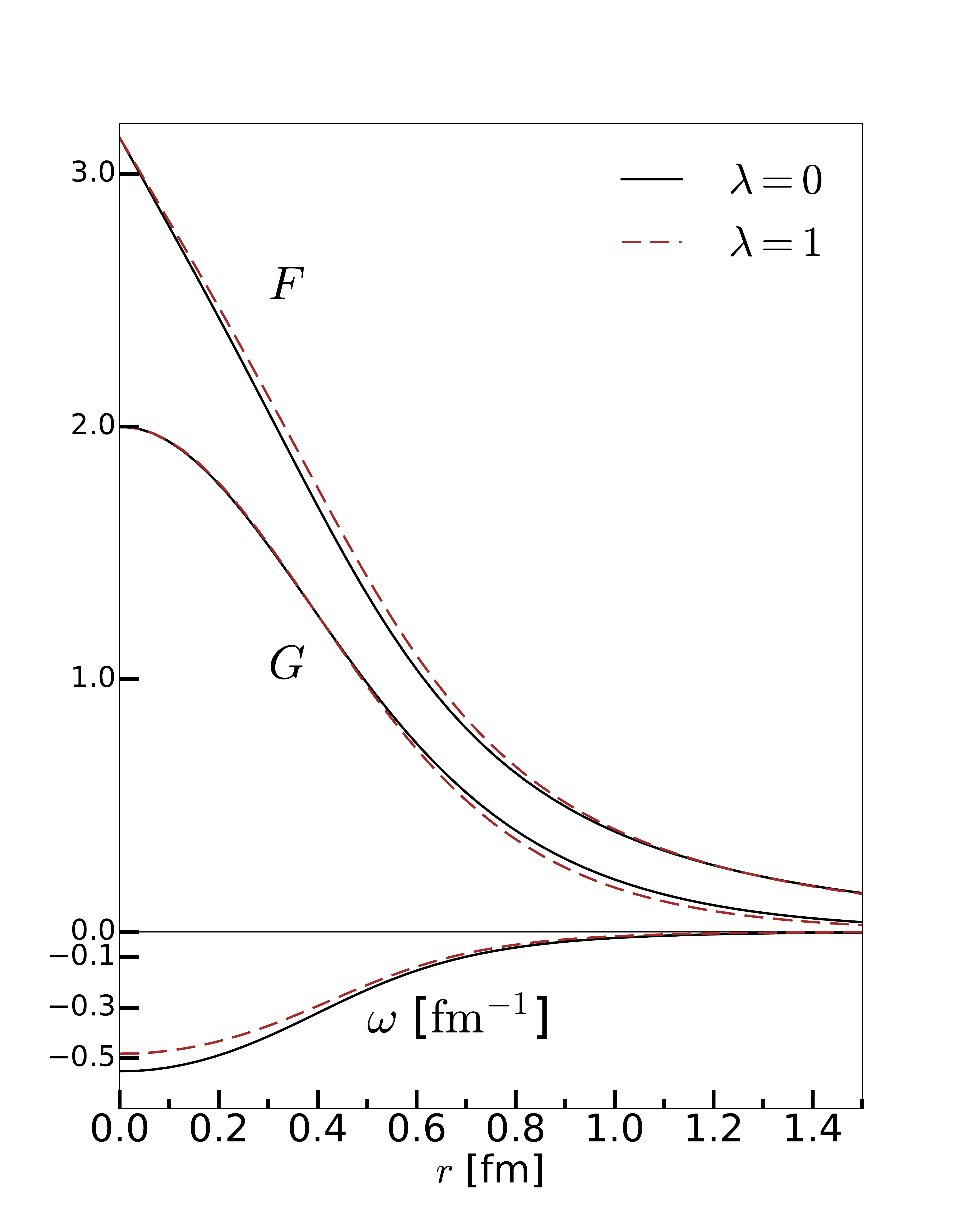}
\caption{The solutions in  free space (solid curves) and in nuclear
  matter at the saturation density $\rho_0$ (dashed curves).}
\label{fig4}
\end{figure}
From Table~\ref{table3} one can see that the solutions are almost
saturated after the third iteration and not so much different from 
the free space ones. This is also seen from
Fig.~\ref{fig4}. Therefore, the factorization of the medium functions
from the model functionals can be considered as a possible
choice for the qualitative discussions.

\vskip 2cm 
\end{document}